\documentclass[floats,aps,prl,twocolumn,showpacs]{revtex4}

\begin{document}
\title{Impurity induced coherent current oscillations in one-dimensional conductors}

\author{S.N. Artemenko}\email[E-mail: ]{art@cplire.ru}
\author{S.V. Remizov} \author{D.S. Shapiro}

\affiliation{Institute for Radio-engineering and Electronics of
Russian Academy of Sciences,
Mokhovaya str. 11-7, Moscow 125009, Russia}

\date{\today}

\begin{abstract}
We study theoretically the electronic transport through a single impurity in a repulsive Luttinger liquid (LL), and find that above a threshold voltage related to a strength of the impurity potential the DC current $\bar I$ is accompanied by coherent oscillations with frequency $f = \bar I/e$. There is an analogy with Josephson junctions: the well-known regime of power-law I-V curves in the LL corresponds to damping of the Josephson current below the critical one, while the oscillatory regime in the LL can be compared with the Josephson oscillations above the critical current.
\end{abstract}

\maketitle

Basic electronic properties of three-dimensional (3D) solids are usually well described within Landau's Fermi-liquid picture where low-energy excitations are quasiparticles that in many respects behave like non-interacting electrons. This is not the case in 1D systems where the usual Fermi liquid picture breaks down. In one dimension single-electron quasiparticles do not exist, the only low energy excitations are charge and spin collective modes. Such a state called the Luttinger liquid (LL) is an alternative to Fermi liquid in 1D (for a review see Ref.~\cite{Giamarchi}). There are different realizations of the 1D electronic systems. The examples are semiconductor-based quantum wires in which dimensionality of the conduction electrons is reduced by dimensional quantization~\cite{Auslaender}, metallic linear chains on Si surfaces~\cite{Yeom}, carbon nanotubes~\cite{carbon}, conducting polymers~\cite{Aleshin}, and quantum Hall effect edge states~\cite{Wen}. There are also evidences that effects related to inter-electronic interaction can be taken into account in terms of LL in strongly anisotropic quasi-1D conductors~\cite{ZZ} where the LL state can be stabilized by defects~\cite{ArtRem} or by formation of the CDW gap induced by electron-phonon coupling~\cite{ArtNat}. The transport properties of the LL are also very different from those of the Fermi liquid. In particular, isolated impurities form effectively large barriers in 1D systems with repulsive inter-electronic interaction and strongly suppress the current which leads to a power-law dependence of conductivity~\cite{KaneFisher,MatveevGlazman,FuruNag}. This effect can be described in terms of tunnelling between different minima of a periodic potential describing interaction of the electronic system with the impurity. The periodic potential is associated with Friedel oscillations induced by impurity. The Hamiltonian describing interaction of the LL with the impurity in spinless LL is expressed in terms of the bosonic (plasmon) displacement field $\hat\Phi (t,x)$ at the impurity position~\cite{Giamarchi}
\begin{equation}
H_{i} = -\frac{e}{\pi} \int dx W_i  \delta (x)\cos 2\hat\Phi, 
\label{impurity}
\end{equation}
where $W_i$ is related to $2k_F$-component of 
the impurity potential. 
And the particle density operator reads~\cite{Giamarchi}
$$
\hat \rho = -\frac{1}{\pi} \frac{\partial \hat\Phi}{\partial x} + \frac{k_F}{\pi }  \cos{ (2k_F x - 2 \hat\Phi)} ,
$$
where the first term describes smooth variations of the particle density and the second one yields a rapidly oscillating part. Fluctuations of the field $\hat\Phi$ make the expectation value of the second term in the free LL equal to zero. However, fluctuations of $\hat\Phi$ are finite at the impurity which results in Friedel oscillations, i. e., in $2k_F$ modulation of charge density decaying with distance as $|x|^{-K_\rho} $, where $K_\rho$  is  the LL parameter measuring the strength of the interaction: $K_\rho <1$ for repulsive and $K_\rho >1$ for attractive interaction, so the larger electronic repulsion the slower decay of the oscillations. As the current operator in the LL reads $\hat I = (e /\pi)\partial_t \hat \Phi$,  the current flow through the impurity implies an increase of $\hat\Phi$ with time, which corresponds to a shift of the Friedel oscillations. 

Power-law I-V curves are induced by tunneling between minima of a washboard potential (\ref{impurity}) slightly inclined by an external bias, so that an increase of the phase by $\pi$ corresponds to a transition of one electron through the impurity~\cite{Giamarchi}. This resembles Josephson junctions where similar fluctuations result in a finite voltage drop at a current below the Josephson critical current, $\bar I < I_c$ (for a review see Ref.~\cite{JosMQT}). However, in superconducting junctions, at $\bar I > I_c$ the Josephson oscillations start, and this corresponds to an increase of the superconducting phase difference with time in the washboard potential which is inclined to an amount exceeding the critical value. Below we show that a similar regime must occur in the LL with an impurity provided the applied voltage exceeds a threshold value corresponding to the slope at which the system can roll out from the minimum of the washboard potential. Above the threshold the current is larger than the tunneling current in the sub-threshold regime of power-law I-V curves, and the current flow is accompanied by oscillations with the washboard frequency $f = \bar I/e$.

For brevity we consider the response of a spinless repulsive LL to an external DC voltage using Tomonaga-Luttinger (TL) model with short range interaction characterized by constant $K_\rho <1$. The short-range interaction describes gated quantum wires where the long-range part of the interaction is screened by 3D gate electrodes. At the end we will discuss essential modifications induced by spin and the long-range Coulomb interaction. 

We start from the Hamiltonian that includes the standard TL Hamiltonian, the impurity part (\ref{impurity}), and the term with an external electric field, $H_E= -\int dx \frac{e}{\pi} E \hat\Phi$,
\begin{equation}
H =  \int dx \frac{\hbar \pi v_{F}}{2} \left[ \hat\Pi^2 +
\frac{1}{\pi^2 K_\rho^2} (\partial_x \hat\Phi)^2\right] + H_{i} + H_E. \label{H0} 
\end{equation} 
Commuting $\hat\Phi$ with the Hamiltonian we derive the equation of motion for the Heisenberg operator 
\begin{equation}
D_0^{-1}\hat\Phi (t,x) = \frac{e}{\pi \hbar}\left[2 W_i  \sin 2 \hat\Phi_0(t) \delta (x) - E \right],
\label{phi}
\end{equation} 
where  $\hat\Phi_0(t) \equiv \hat\Phi(t, x=0)$, and the operator in the left-hand side is the inverse free bosonic Green's function of density fluctuations. For the standard LL it reads 
$$ 
D_0^{-1} =  \left(  v^2 \partial^2_{x} - \partial^2_{t}\right)/(\pi v_F),
$$
where $v = v_F / K_\rho$ is the velocity of plasmons. This operator does not contain damping. If one takes into account coupling of electrons to a dissipative bosonic bath (to phonons or to density fluctuations in a metallic gate) then a finite damping appears, and the Fourier transformed retarded Green's function acquires the form~\cite{Guinea}
\begin{equation}
D_0^R = \frac{\pi v_F}{\omega^2 + i \omega \nu - \omega^{2}_{q}} , \quad \omega^{2}_{q} = q^2v^2. 
\label{D0}
\end{equation} 
As it was shown recently the damping suppresses soliton-like fluctuations and reduces power-law conductance to an exponential one~\cite{Natter}. 

Using Eq.~(\ref{phi}) with proper boundary conditions at the contacts one can express $\hat\Phi(t,x)$ in terms of its value at the impurity site. We apply boundary conditions~\cite{Grabert} derived for a wire adiabatically connected to ideal Fermi-liquid reservoirs at $x= \pm L/2$ with voltage difference $V$. Then by means of the Fourier transformation we finally obtain the equation of motion for the operator $\hat\Phi_0(t)$ at the impurity position.
\begin{eqnarray}
&&
\partial_t \hat\Phi_0 (t) +  \frac{e}{\hbar}\int_0^\infty dt_1  Z(t-t_1) \sin 2 \hat\Phi_0 (t_1) = \frac{\pi}{e} \bar I,
\label{phi-t0}  \\
&&
Z (t) =  \int \frac{d\omega }{2\pi} e^{- i\omega t} \frac{W_i \tilde K_\rho(1 - i  \tilde K_\rho \tan \frac{q_0L}{2})}{(\tilde K_\rho - i \tan \frac{q_0L}{2})} \nonumber
\\
&&
q_0 = \frac{\sqrt{\omega^2+ i\omega \nu}}{v}, \quad \tilde K_\rho = K_\rho \sqrt{\frac{\omega}{\omega + i\nu}},
\nonumber
\end{eqnarray} 
where $\bar I = e^2 E v_F/\pi \hbar \nu$ is the time averaged macroscopic current in the channel.
Eq.~(\ref{phi-t0}) is supplemented by relation between time averaged values
\begin{equation}
\bar I = G (V-V_{i}), \quad G^{-1}=G_0^{-1}+G_{\nu}^{-1},\quad G_{\nu} =\frac{e^2 v_F}{\pi \hbar \nu L},
\label{I0}
\end{equation} 
where $V_{i} =2W_i \langle \sin 2 \hat\Phi_0(t) \rangle_t$ is the DC component of the voltage drop at the impurity, $G_0 = e^2/h$ is the conductance quantum per spin orientation, and $G_{\nu}$ is the conductance related to damping in the clean part of the wire.

As we want to concentrate on conduction through the impurity and not on the problem of contacts, we will consider the case of small enough damping, $\nu \ll \omega$, and a long channel, $L \gg l_\nu = v/\nu$. This allows to neglect reflections of current pulses generated by the impurity from the contacts and to substitute $\tan \frac{q_0L}{2} \to i$. Then the kernel $Z$ simplifies
\begin{equation}
Z (t) = W_i K_\rho \int  \frac{d\omega }{2\pi}e^{- i\omega t} \sqrt{\frac{\omega}{\omega + i \nu}} ,
\label{Z2}
\end{equation} 
which at $\nu =0$ gives $Z (t) \to W_i K_\rho \delta(t)$. Note, however, that at any small but finite damping $\int Z(t) dt =0$, so the damping cannot be neglected at $\omega =0$.

Remember that $\hat \Phi_0 (t)$ in (\ref{phi-t0}) is an operator, so it is not easy to solve this non-linear equation. So we extract the expectation value $\Phi_0(t)=\langle \hat\Phi_0 \rangle$, thus $\hat\Phi_0 \equiv \Phi_0 + \delta \hat\Phi$, where $\langle \delta \hat\Phi \rangle =0$, and $\Phi_0(t)$ satisfies the equation
\begin{equation}
\partial_t \Phi_0 (t) +  \frac{e}{\hbar}\! \int_0^\infty \!\!\! dt_1 Z(t-t_1) \langle \cos 2\delta \hat\Phi (t_1) \rangle \! \sin 2 \Phi_0 (t_1) = \frac{\pi}{e} \bar I .
\label{phi0}
\end{equation} 
To solve this equation one must calculate $\langle \cos 2\delta \hat\Phi \rangle$ first. In calculation of this expectation value we will ignore the soliton-like fluctuations that are responsible for sub-threshold tunneling, but take into account Gaussian fluctuations that substantially reduce $\langle \cos 2\delta \hat\Phi \rangle $. These fluctuations can be treated by means of the self-consistent harmonic approximation. 

But before treating the general case we discuss solution in the simple limit of strong inter-electron interaction ($K_\rho \to 0$) when Eq.~(\ref{phi0}) can be easily solved analytically. In this limit fluctuations of the displacement field at the impurity are small, $\langle \cos 2\delta \hat\Phi \rangle \to 1$, and $\hat \Phi_0 (t)$ can be treated as $c$-number. When $V \leq V_T = 2W_i$ we find from Eqs.~(\ref{I0}) and (\ref{phi0}) a stationary solution $2\Phi_0 = \arcsin (V/V_T) $ with zero current $I =0$ but with non-zero voltage drop over impurity. Note that we obtain that the current is zero in the sub-threshold region because we neglected solitonic fluctuations resulting in power-law I-V curves. 

At $V > V_T$ the solution is oscillatory with fundamental frequency  $f = \bar I /e $
\begin{equation}
\partial_t  \Phi_0 (t) = \frac{\pi \bar I^2/e}{ \sqrt{\bar I^2 + I_0^2} + I_0 \sin (2\pi \bar I /e) t},
\label{dphi0}
\end{equation}
where $I_0 = G_0 V_T K_\rho$. Eq.~(\ref{dphi0}) determines the current at the impurity site, $I (t, x=0) = e \partial_t  \Phi_0 (t) / \pi $. Current at the clean part of the channel calculated from Eq.~(\ref{phi}) is equal to $I(t,x) = e \partial_t \Phi_0 (t-|x|/v)/\pi$ at $|x| \ll l_\nu$, and $I(t,x) = \bar I$ at large distances from the impurity. 

For the DC voltage drop at the impurity we obtain
\begin{equation}
V_{i} = V_T \frac{\sqrt{\bar I^2 + I_0^2}- \bar I}{I_0}.
\label{s}
\end{equation}
Thus from (\ref{I0}), (\ref{dphi0}) and (\ref{s}) we see that the oscillatory regime starts at $V > V_T$. The current at the impurity consists of narrow  pulses of height $2I_0$ at $\bar I \ll I_0$ and transforms into the Ohmic current accompanied by harmonic oscillations of amplitude $I_0$ at $\bar I \gg I_0$.

Now we consider the case of finite values of $K_\rho$.
Mean square fluctuations $\langle \delta \hat\Phi^2 \rangle$ can be calculated from Keldysh Green's function $D^K = -i \langle \{ \delta \hat\Phi(t),\delta \hat\Phi(t')\}_+  \rangle$ at $t=t'$. This function can be expressed via retarded and advanced Green's functions 
$$D^{R(A)} = \pm i\theta [\pm(t-t')]  \langle [ \delta \hat\Phi(t),\delta \hat\Phi(t')]_-  \rangle$$ by relation $D^K (t,t')=D^R (t,t')f(t')-f(t)D^A (t,t')$, where Fourier transform of $f$ is related to the distribution function of bosonic excitations $N (\omega)$, $f(\omega) = 1 + 2N (\omega)$. In the equilibrium state $N (\omega)$ is the Planck distribution. At low temperatures, smaller than all characteristic energies of the system, one can neglect contribution of thermally excited excitations, then $f(\omega) = {\rm sign} (\omega)$ and  $D^K$ can be expressed via the retarded and advanced functions. Note that acting in this way we neglect the effect of non-equilibrium distribution of bosonic excitations.

Now we derive equations of motion for the retarded and advanced Greens functions of fluctuations. This can be done in a standard way multiplying Eq.~\ref{phi} by $\hat\Phi (t',x')$ from the left and from the right, and combining them in order to obtain corresponding Green's function after averaging. Then using the Fourier transformation we express Greens functions at the impurity site and get rid of the coordinate dependence of Green's function. After that by means of (\ref{phi0}) we subtract expectation values and use the self-consistent harmonic approximation
$$
\sin 2 \delta \hat\Phi \to  \langle \cos 2\delta \hat\Phi \rangle 2 \delta \hat\Phi = e^{-2 \langle \delta \hat\Phi^2 \rangle} 2 \delta \hat\Phi,
$$
and arrive, finally, to close equations of motion for $D^{R(A)}$. As we will need this equations at frequencies larger than the small damping constant we neglect $\nu$ and employ $Z (t) = W_i K_\rho \delta(t-t_1)$. Then the equation for $D^{R}$ reads
\begin{eqnarray}
&&
\partial_t D^R(t,t') +  \frac{2e}{\hbar} W_i K_\rho C(t)   D^R(t,t') = -\frac{\pi K_\rho}{2} \delta (t,t'), \nonumber \\
&&
 C(t) \equiv  \cos 2 \Phi (t_1) \langle \cos 2\delta \hat\Phi (t_1) \rangle.
\label{DR}
\end{eqnarray} 
This equation and the similar equation for the advanced Green's function can be easily solved analytically. The solutions are
\begin{equation}
D^{R(A)} = - \frac{\pi K_\rho}{2} \theta [\pm(t-t')]e^{ \mp \frac{2e}{\hbar}W_i K_\rho \int^{t}_{t'} C(t_1) dt_1 }.
\label{DRA}
\end{equation}
This gives us for $\langle \delta \hat\Phi^2 \rangle = \frac{i}{2}D^{K}(t,t)$
\begin{equation}
\langle \delta \hat\Phi^2 \rangle = \frac{K_\rho}{2} \int^{\infty}_{0} \frac{dt_1}{t_1} e^{ - \frac{2e}{\hbar}W_i K_\rho \int^{t_1}_{0} C(t-t_2) dt_2 }  .
\label{DK}
\end{equation}
Eq.~(\ref{DK}) must be solved self-consistently with (\ref{phi0}) and (\ref{I0}) in order to find $\langle \delta \hat\Phi^2 \rangle$ as function of $\cos 2 \Phi_0$. First, we calculate the threshold voltage. In the sub-threshold region $C$ does not depend on time. Then calculating integrals in (\ref{DK}) and using definition of $C$ (\ref{DR}) we obtain 
\begin{equation}
\langle \delta \hat\Phi^2 \rangle =\frac{K_\rho}{2(1-K_\rho)} \ln \frac{\Lambda}{2K_\rho eW_i \cos 2 \Phi_0} ,
\label{DPhi}
\end{equation} 
where $\Lambda \sim p_F v $ is a large cut-off energy. Note that in accordance with our previous statement we found that fluctuations vanish at $K_\rho \to 0$.
Substituting (\ref{DPhi}) in (\ref{I0}) we obtain equation for $\Phi_0$ in the sub-threshold regime
\begin{equation}
2 W_i  \left( \frac{2K_\rho eW_i \cos 2 \Phi_0 }{\Lambda}\right)^{\frac{K_\rho}{1-K_\rho}} \sin 2 \Phi_0 =  V.
\label{phi0st}
\end{equation}
It has solution at $V < V_T$ with
\begin{equation}
V_T = 2W_i  \left( \frac{2K_\rho^{3/2} eW_i}{\Lambda}\right)^{\frac{K_\rho}{1-K_\rho}}\sqrt{1-K_\rho}.
\label{UT}
\end{equation}
We see from (\ref{UT}) that the finite threshold voltage exists provided $K_\rho < 1$, and this is in accordance with the condition that the impurity is a relevant perturbation in case of repulsive inter-electronic interaction. 

It is not simple to find $\langle \delta \hat\Phi^2 \rangle$ analytically in a general time-dependent case, but this can be done easily when the voltage slightly exceeds the threshold value and the DC part of the current is small, $\bar I \ll I_0 $. Then $\Phi_0$ increases with time in a step-like manner spending the most part of the oscillation period $e/\bar{I}$ near the value corresponding to maxima of $C(t)$ and passes rapidly other values of $C$ during short time interval $\sim 1/\sqrt{\bar I I_0}$ that gives small contribution to the period. Therefore, in the most part of the period one can use (\ref{DPhi})  with time dependent value $\Phi_0(t)$, and self-consistent solution of Eqs.~(\ref{DK}) and (\ref{phi0}) in this limit yields equations similar to (\ref{dphi0}) and (\ref{s}) with the value of the threshold voltage $V_T$  (\ref{UT}).

So far we considered the spinless LL. In the spinful LL impurities partly violate the spin-charge separation, and the impurity Hamiltonian contains the spin phase field~\cite{Giamarchi}. This leads to modification of the results, in particular, of the threshold voltage. For a spin-independent electronic repulsion Eq.~(\ref{UT}) is substituted by
\begin{equation}
V_T = \sqrt{2}W_i  \left( \frac{2 eV}{\Lambda}\right)^{\frac{1+K_\rho}{1-K_\rho}}K_\rho^{\frac{1+K_\rho}{3K_\rho}}\sqrt{2(1-K_\rho)}.
\nonumber 
\end{equation}

Now we discuss the case of the long-range Coulomb repulsion. It can be taken into account in terms of momentum dependence of $K_\rho$~\cite{Schulz}
$$
K_\rho^{-2} = 1+ a^2 {\rm K_0} (qd), \quad a^2 = \frac{4e^2}{\pi \hbar v_F \epsilon},
$$
where the McDonald function originates from the Fourier transformation of the Coulomb potential in a wire of diameter $d$, $\epsilon$ is a background dielectric constant. In case of the long-range interaction the impurity is a relevant perturbation at all values of the parameter $a$ describing the strength of the repulsion. This is evident from the flow equation demonstrating that the impurity potential scales to infinity under renormalization
$$
\frac{dW_i}{dl}=
\left[1-\left(1 + a^2\ln{\frac{\hbar \omega_d}{\Lambda}}\right)^{-1/2}\right]W_i, \quad \omega_d = \frac{2a v_F}{d}.
$$
The McDonald function is substituted here by its limiting logarithmic form.

The momentum dependence of the interaction parameter modifies equations (\ref{phi-t0}) and (\ref{phi0}). With the logarithmic accuracy and for the case of negligible damping we find that $Z(t)$ (\ref{Z2}) must be substituted for 
\begin{equation}
Z (t) = W_i K_\rho \int  \frac{d\omega }{2\pi}e^{- i\omega t} \sqrt{\frac{\omega}{(\omega + i \nu)(1 + a^2 \ln \frac{\omega_d}{|\omega| })}} .
\nonumber
\end{equation}
Though it is not simple to solve such equations analytically we find that they do not result in important qualitative difference from the case of the short-range interaction with constant $K_\rho$. One of the most important distinctions is that suppression of $\langle \cos 2\delta \hat\Phi \rangle$ by fluctuations is smaller than in the TL model, and this results in the smaller effect of fluctuations on the threshold voltage. Calculating again $\langle\delta \hat\Phi^2 \rangle$ by means of Keldysh Green's function we find in the sub-threshold regime
\begin{equation}
\langle\delta \hat\Phi^2 \rangle= \frac{1}{2}\int_0^\infty 
\frac{d\omega}{\omega\sqrt{1+a^2\ln \frac{\omega_d}{\omega }} + 2 (e/\hbar) W_i C}. \label{flrow} 
\end{equation}
Assuming a moderate strength of the Coulomb repulsion when $a$ is of the order one, and taking into account that the argument of the logarithm in (\ref{flrow}) is large, we estimate $\langle\delta \hat\Phi^2 \rangle$ with the logarithmic accuracy. Then we obtain
$$
\langle\delta \hat\Phi^2 \rangle=\frac{1}{a}\sqrt{\ln\frac{e}{\epsilon d W_i C}.} \label{f1} 
$$
Here we neglected the contribution from high-energy cut-off, assuming that $\ln \frac{\omega_d}{\Lambda} \sim \ln \frac{a}{k_F d}$ is not a large value.
This gives an estimate for the threshold field 
$$ V_T = 2 W_i \exp\left (-\frac{2}{a}\sqrt{\ln\frac{e}{\epsilon d W_i}}\right ),
$$
where, again, the argument of the square root in the exponential function is given with the logarithmic accuracy.

In conclusion, we found that above the threshold voltage $V_T$ the current through an impurity generates coherent oscillations with the fundamental frequency $f = \bar I/e$. We considered the case of the DC applied voltage. If in addition there is also an AC component of the applied voltage with frequency $f_{0}$ then an analog of the Shapiro steps observed in Josephson junctions will appear on the I-V curves. In our problem these are the steps of a constant voltage, the fundamental step being located at the current value $\bar I = e f_0$. Characteristic frequencies of the oscillations induced by the DC voltage are determined by the strength of the impurity potential. In semiconducting quantum wires typical values of the impurity potential (say, from the shallow impurities) can be of the order of several meV, and depending on the strength of the electronic repulsion the frequency may fall into the gigahertz or terahertz frequency regions. Direct application of our results to real systems is limited by voltages smaller than distances to other electronic subbands. The results can be modified also due to different coupling of the 1D system to 3D environment.

We thank K.E. Nagaev and V.A. Sablikov for useful discussions. The work was supported
by Russian Foundation for Basic Research. A part of the research was performed in the frame of the
CNRS-RAS-RFBR Associated European Laboratory ``Physical properties
of coherent electronic states in condensed matter'' between Institut N\'eel, CNRS
and IRE RAS.

\end{document}